\documentclass[twocolumn]{aastex701}
\usepackage{amsmath}
\usepackage{xcolor}

\submitjournal{ApJL}
\accepted{2026 August 11}
\begin{document}

\title{Statistics of Solar Filament Mass based on CHASE Sun-as-a-star Spectroscopic Observations}

\author[orcid=0009-0001-8176-5747]{T. Y. Xie}
\affiliation{School of Astronomy and Space Science, Nanjing University, Nanjing 210023, China}
\affiliation{Key Laboratory of Modern Astronomy and Astrophysics, Ministry of Education, Nanjing 210023, China}
\email{tyxie0304@163.com}

\author[orcid=0009-0003-8956-547X]{Z. H. Zhao}
\affiliation{School of Astronomy and Space Science, Nanjing University, Nanjing 210023, China}
\affiliation{Key Laboratory of Modern Astronomy and Astrophysics, Ministry of Education, Nanjing 210023, China}
\affiliation{Department of Astronomy, Graduate School of Science, Kyoto University, Kyoto, Japan}
\email{}

\author[orcid=0000-0003-2837-7136]{X. Cheng}
\affiliation{School of Astronomy and Space Science, Nanjing University, Nanjing 210023, China}
\affiliation{Key Laboratory of Modern Astronomy and Astrophysics, Ministry of Education, Nanjing 210023, China}
\email[show]{xincheng@nju.edu.cn}
\correspondingauthor{X. Cheng}

\author[orcid=0000-0003-1220-1582]{Y. H. Chen}
\affiliation{School of Astronomy and Space Science, Nanjing University, Nanjing 210023, China}
\affiliation{Key Laboratory of Modern Astronomy and Astrophysics, Ministry of Education, Nanjing 210023, China}
\affiliation{Leibniz Institute for Astrophysics, Potsdam, Germany}
\email{}

\author[orcid=0009-0006-5180-8052]{Z. Zheng}
\affiliation{School of Astronomy and Space Science, Nanjing University, Nanjing 210023, China}
\affiliation{Key Laboratory of Modern Astronomy and Astrophysics, Ministry of Education, Nanjing 210023, China}
\email{zz_nju@smail.nju.edu.cn}

\author[orcid=0000-0002-9264-6698]{Q. Hao}
\affiliation{School of Astronomy and Space Science, Nanjing University, Nanjing 210023, China}
\affiliation{Key Laboratory of Modern Astronomy and Astrophysics, Ministry of Education, Nanjing 210023, China}
\email{haoqi@nju.edu.cn}

\author[orcid=0000-0001-7693-4908]{C. Li}
\affiliation{School of Astronomy and Space Science, Nanjing University, Nanjing 210023, China}
\affiliation{Key Laboratory of Modern Astronomy and Astrophysics, Ministry of Education, Nanjing 210023, China}
\email{}

\author[orcid=0000-0002-4978-4972]{M. D. Ding}
\affiliation{School of Astronomy and Space Science, Nanjing University, Nanjing 210023, China}
\affiliation{Key Laboratory of Modern Astronomy and Astrophysics, Ministry of Education, Nanjing 210023, China}
\email{}

%\collaboration{all}{The Terra Mater collaboration}

\begin{abstract}
Filaments are cool and dense plasmas suspended in the hot corona of the Sun and other stars. Accurately estimating their masses is of great significance for understanding subsequent eruptions and induced space weather effects, but it remains hindered by their intrinsic geometric uncertainties, particularly in spatially unresolved stellar observations. To test and calibrate the methods for estimating the masses of stellar filaments, we conduct a statistical Sun-as-a-star analysis of solar filaments, utilizing full-disk H$\alpha$ spectroscopic observations from the Chinese H$\alpha$ Solar Explorer (CHASE). A total of 1346 filaments, covering a period from January 2024 to October 2025, are identified via a machine-learning segmentation model. We construct their virtual sun-as-a-star spectra by spatially integrating the filament regions and then obtain their optical parameters by cloud-model fitting. Upon correcting projection effects, we establish a representative three-dimensional morphological scaling of length, apparent width, and line-of-sight depth ($L:W_{\rm app}:D_{\rm LOS} \approx 4.5:1:1.7$), with a median filament depth of about $8000 \ \rm{km}$. Interestingly, the Sun-as-a-star estimated mass shows high consistency with the resolved intrinsic mass across the full sample, with a log-space regression slope of 1.07. As the first large-sample Sun-as-a-star study of solar filaments, our results provide empirical constraints on filament geometries and masses, offering a critical reference for estimating stellar filament masses based on H$\alpha$ spectroscopy.

\end{abstract}


\section{Introduction} 
\label{sec:intro}

Solar filaments are cool and dense plasmas suspended in the hot, tenuous solar corona, with temperatures of $T \sim 10^4 \ \rm K$ and electron densities of $n_e \sim 10^9-10^{11}\ \mathrm{cm^{-3}}$ \citep[e.g.,][]{labrossePhysicsSolarProminences2010, mackayPhysicsSolarProminences2010a}. As the Sun rotates, filaments will appear as bright prominences once located above the visible solar limb. Typically forming above magnetic polarity inversion lines (PILs) \citep{gaizauskasFormationSolarFilament1997}, they are supported against gravity by the tension of magnetic flux ropes or sheared arcades \citep[e.g.,][]{ballegooijenFormationEruptionSolar1989, martinConditionsFormationMaintenance1998, Cheng_2017, patsourakosDecodingPreEruptiveMagnetic2020}. Observations have revealed abundant fine dynamics within filaments, including persistent mass circulation in the form of upflows and downflows \citep{bergerHinodeSOTObservations2008}, underscoring their role as dynamically active structures rather than static plasma reservoirs. Owing to their close coupling with the coronal magnetic topology, filaments act as sensitive diagnostics of the magnetic field configuration and constitute essential elements of the chromosphere–corona mass exchange cycle \citep{parentiSolarProminencesObservations2014}.

The mass of the filaments plays a crucial role in regulating the equilibrium of the corresponding magnetic systems: as filament material gradually drains, the reduced loading allows the systems to rise until the equilibrium is lost, thereby resulting in solar eruptions—a physical mechanism originally proposed by \citet{lowSolarActivityCorona1996} and recently demonstrated via three-dimensional full-MHD simulations \citep{Xing_2025}. Such eruptions frequently develop into coronal mass ejections (CMEs) \citep[e.g.,][]{gopalswamyProminenceEruptionsCoronal2003, mccauleyProminenceFilamentEruptions2015}, which are a primary driver of disastrous space weather \citep[e.g.,][]{webbCoronalMassEjections2012a}. The initial mass of erupting filaments, therefore, constitutes a key parameter in characterizing the properties of CMEs and their potential space weather impacts, underscoring the need for accurate filament mass measurements.

Over the past few decades, various methods have been developed to estimate filament mass based on multi-wavelength data, including EUV opacity measurements \citep[e.g.,][]{gilbertCOMPARINGSPATIALDISTRIBUTIONS2010} and H$\alpha$ spectral inversions \citep[e.g.,][]{heinzelTheoreticalCorrelationsProminence1994, heinzelRadiativeTransferSolar2015d}. Although individual case studies have achieved reasonable accuracy in determining the column density of filaments, a fundamental limitation persists in all methods: the optical depth and the geometric depth of filaments along the line of sight cannot be independently constrained, preventing an unambiguous determination of the column density and, hence, the total mass. This geometric ambiguity is particularly pronounced in on-disk observations, where projection effects obscure the true spatial scale of filaments. Critically considered, the magnitude and systematic behavior of this geometric degeneracy have never been quantified across a large sample, representing a significant gap in our understanding of filament mass budgets.

In the broader context of stellar physics, superflares on solar-type stars have emerged as an active area of research with the advent of time-domain astronomy \citep{maeharaSuperflaresSolartypeStars2012}, and associated stellar CMEs — predominantly driven by filament eruptions — represent a significant channel of mass loss in young, active stars \citep[e.g.,][]{ostenCONNECTINGFLARESTRANSIENT2015}. Candidate stellar CME signatures have been detected via blueshifted absorption or emission features in Balmer lines \citep{vidaInvestigatingMagneticActivity2016, leitzingerCensusCoronalMass2020b, luExtremeStellarProminence2025a} and coronal dimmings at EUV and X-ray bands \citep{veronigIndicationsStellarCoronal2021a}. Notably, \citet{namekataProbableDetectionEruptive2022} presented a mass estimation from spatially unresolved H$\alpha$ spectra on a solar-type star, illustrating the potential of spectroscopic diagnostics for characterizing stellar filament eruptions. They implicitly assume that stellar filaments share the fundamental physical and geometric properties of their solar counterparts. However, this assumption has never been empirically validated against spatially resolved solar observations — which is precisely the gap that this work addresses.

To address these challenges, we present a statistical analysis of 1346 solar filaments from full-disk H$\alpha$ hyperspectral observations by the Chinese H$\alpha$ Solar Explorer (CHASE). Filaments are detected automatically using a U-Net deep learning pipeline \citep{illarionovMachinelearningApproachIdentification2020, zhengDevelopingAutomatedDetection2024a}. Our study delivers two key results. First, we provide the first population-level empirical constraints on the aspect ratios of solar filaments, characterized by their length-to-width-to-depth ratio. Second, we construct Sun-as-a-star H$\alpha$ spectra to simulate spatially unresolved stellar observations and validate the resulting disk-integrated mass estimates against the spatially resolved filament mass, establishing a solar-calibrated scaling relation for stellar filament mass inference.

\section{Data and Observations}
\label{sec:data}

The data utilized in this study are from the Chinese H$\alpha$ Solar Explorer (CHASE). As China's first space-based solar telescope dedicated to H$\alpha$ imaging spectroscopy, CHASE is designed to capture the dynamical processes of the chromosphere with high temporal and spatial resolutions \citep{2022CHASELi, qiuCalibrationProceduresCHASE2022b}. Its primary payload, the H$\alpha$ Imaging Spectrograph (HIS), provides full-disk scanning capabilities, acquiring solar spectra simultaneously at H$\alpha$ ($6559.7-6565.9 \ \mathrm{\AA}$) and \ion{Fe}{1} ($6567.8$--$6570.6 \ \mathrm{\AA}$) wavebands, with a spatial resolution of $0.52 \ \mathrm{arcsec}$ and a spectral sampling of $0.024 \ \mathrm{\AA}$. Operating beyond Earth's atmosphere, CHASE provides continuous, seeing-free monitoring from the photosphere to the chromosphere, capturing the fine structures and dynamics of chromospheric features.

For this statistical study, we employ the full-disk intensity images at the H$\alpha$ line center ($6562.8 \ \mathrm{\AA}$). To construct a representative dataset, we select CHASE observations from January 1, 2024 to October 23, 2025 with a uniform cadence of 12\,days, yielding a total of 54 observing epochs.

Filament identification is performed using the automated segmentation pipeline developed by \citet{zhengDevelopingAutomatedDetection2024a}. This deep learning model, built upon the U-Net architecture \citep[e.g.,][]{ronnebergerUNetConvolutionalNetworks2015, ibtehazMultiResUNetRethinkingUNet2020, siddiqueUNetVariantsMedical2021, azadMedicalImageSegmentation2024} and specifically trained on \textit{CHASE}/HIS spectral imagery, is selected because of its proven robustness in handling the instrument-specific characteristics of the data. The model takes the calibrated full-disk H$\alpha$ line-center intensity maps as input and outputs a binary segmentation mask in which filament structures are effectively isolated from the quiet Sun background and active regions. No additional manual interventions are applied to the segmentation results, ensuring the objectivity and reproducibility of the statistical sample.

\begin{figure*}[t!]
    \centering
    \includegraphics[width=\textwidth]{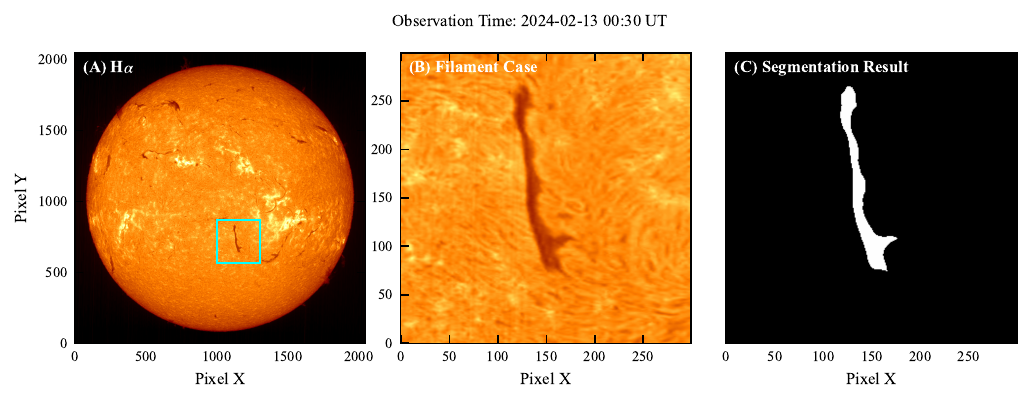}
    \caption{Filament identification workflow. (A) Full-disk H$\alpha$ line-center image observed by CHASE at 00:30~UT on 2024 February 13. The cyan rectangle marks the zoomed region shown in Panels~(B) and~(C). (B) H$\alpha$ line-center image of a representative filament. (C) Corresponding U-Net segmentation result, with the identified filament region shown in white.}
    \label{fig:identification}
\end{figure*}

In total, 1346 filament instances are identified across the 54 selected observing epochs. The sample is dominated by quiescent filaments, with active region filaments comprising only a small fraction of the total. Figure~\ref{fig:identification} illustrates a representative example of the data processing and identification workflow. 

\section{Methodology}
\label{sec:methods}

The primary objective of this study is to build the solar--stellar filament mass connection by comparing spatially resolved filament mass estimates with those derived from disk-integrated, Sun-as-a-star spectra. Our analysis consists of two main stages: spectral inversion via the cloud model (see Section~\ref{sec:cloud_model}), followed by mass estimation under two complementary observational assumptions (see Section~\ref{sec:mass_calc}).

\subsection{Spectral Inversion via Cloud Model}
\label{sec:cloud_model}

To retrieve the physical properties of the filament plasma from the H$\alpha$ spectra, we employ the classical cloud model \citep{meinDifferentialCloudModels1988}, which treats the filament as a slab of plasma suspended above the solar surface that modifies the background radiation through absorption and emission.

For a given pixel at position $(x, y)$, the observed intensity profile $I(\lambda)$ is expressed as the result of the photospheric emission crossing a cool cloud \citep[e.g.,][]{Zirin1988,Foukal1990,Aschwanden2004}:
\begin{equation}
    I(\lambda) = I_0(\lambda)\,e^{-\tau(\lambda)} + S\left[1 - e^{-\tau(\lambda)}\right],
    \label{eq:cloud_model}
\end{equation}
where $I_0(\lambda)$ is the background quiet-Sun intensity profile, $S$ is the source function (assumed constant within the cloud), and $\tau(\lambda)$ is the wavelength-dependent optical depth. Assuming a Gaussian absorption profile, the optical depth is parameterized as 
\begin{equation}
    \tau(\lambda) = \tau_0 \exp \left[ -\left( 
    \frac{\lambda - \lambda_0 - \Delta\lambda_D}{W} \right)^2 \right],
    \label{eq:tau}
\end{equation}
where $\tau_0$ is the line-center optical depth, $\lambda_0 = 6562.8 \ \mathrm{\AA}$ is the rest wavelength of H$\alpha$, $\Delta\lambda_D = \lambda_0 v_D / c$ is the wavelength shift corresponding to the line-of-sight Doppler velocity $v_D$, and $W$ is the Doppler width.

For each pixel within the identified filament region, Equation~\eqref{eq:cloud_model} is fitted to the observed spectral profile to simultaneously derive the four cloud model parameters: $\tau_0$, $S$, $v_D$, and $W$. This inversion yields two-dimensional maps of the physical properties across each filament.

\subsection{Sun-as-a-star Spectral Synthesis}
\label{sec:sun_as_star}
To simulate the observational signal of a solar filament as viewed from a distant star with no spatial resolution, we construct virtual Sun-as-a-star difference spectra following a modified version of the methodology described in \citet{otsuSunasastarAnalysesVarious2022}. Since this study focuses on the statistical properties of quiescent filaments rather than their temporal evolution, we treat each identified filament as a static snapshot and define a spatial reference background rather than a temporal one. The spatially averaged spectral intensity over a region $\mathrm{A}$ is defined as
\begin{equation}
    f(\lambda, \mathrm{A}) = \frac{1}{N(\mathrm{A})} 
    \int_{\mathrm{A}} I(\lambda, x, y) \, \mathrm{d}x\, \mathrm{d}y,
\end{equation}
where $N(\mathrm{A})$ is the total number of pixels in the region. To facilitate comparison and mitigate local continuum variations, the spectra are normalized by the local continuum intensity at a reference wavelength $\lambda_{\mathrm{cont}}$ adjacent to the H$\alpha$ line:
\begin{equation}
    F(\lambda, \mathrm{A}) = \frac{f(\lambda, \mathrm{A})}{f(\lambda_{\mathrm{cont}}, \mathrm{A})}.
\end{equation}

The Target Region (TR) consists of the pixels belonging to the filament body, whereas the Background (BG) is defined as the region of 10 pixels distributed outside the filament boundary.

The Sun-as-a-star difference spectrum, $\Delta S_{\mathrm{TR}}(\lambda)$, representing the net absorption signature of the filament on the integrated stellar disk, is obtained by subtracting the normalized background spectrum from the filament spectrum and scaling it by the fractional disk area:
\begin{equation}
    \Delta S_{\mathrm{TR}}(\lambda) = 
    \frac{F(\lambda, \mathrm{TR}) - F(\lambda, \mathrm{BG})}
         {F(\lambda_{\mathrm{cont}}, \mathrm{FD})} 
    \times \frac{N(\mathrm{TR})}{N(\mathrm{FD})},
    \label{eq:spatial_diff}
\end{equation}
where $N(\mathrm{FD})$ is the total number of pixels covering the full solar disk. This quantity simulates the spectral signature that a solar filament would produce in a disk-integrated stellar observation. We note that this formulation implicitly assumes a single filament on the disk; the impact of multiple filaments and the sensitivity of the results to the choice of $\mathrm{TR}/\mathrm{BG}$ size represent important caveats that lie beyond the scope of the present study and are deferred to future work.

The equivalent width of the filament absorption feature is then obtained by integrating this difference spectrum over the H$\alpha$ line within a window of $\Delta\lambda = 0.72$~\AA:
\begin{equation}
    \Delta \mathrm{EW}_{\mathrm{obs}} = 
    \int_{\lambda_0 - \Delta\lambda}^{\lambda_0 + \Delta\lambda} 
    \Delta S_{\mathrm{TR}}(\lambda) \, \mathrm{d}\lambda.
\end{equation}

\subsection{Mass Estimation}
\label{sec:mass_calc}

We derive the filament masses using two complementary approaches that exploit different levels of spatial information, enabling a direct assessment of the accuracy of stellar mass estimation methods against the spatially resolved filament mass.

\subsubsection{Spatially Resolved Mass}
\label{sec:resolved_mass}

The spatially resolved mass is derived following \citet{tsiropoulaDeterminationPhysicalParameters1997}.The total hydrogen number density $N_\mathrm{H}$ is computed from the cloud-model parameters via
\begin{equation}
    N_\mathrm{H} = \frac{3.2\times10^{8}}{i(T,\,p)}\,\sqrt{N_2},
    \label{eq:NH}
\end{equation}
where $N_2$ is the $n{=}2$ level population density derived from the optical depth and Doppler width \citep{polandHydrogenIonizationN21971, yakovkinProminenceRadiationTheory1975},and $i(T,p)$ is the hydrogen ionization degree tabulated as a function of temperature and gas pressure in \citet{heinzelRadiativeTransferSolar2015d}. Based on semi-empirical prominence models of \citet{zhangSemiempiricalModelsQuiescent1987} and observational constraints of \citet{tsiropoulaDeterminationPhysicalParameters1997}, we adopt $T = 6500$--$8000$\,K and $p = 0.15$--$0.20$\,dyn\,cm$^{-2}$, which yield $i = 0.24$--$0.40$ and a corresponding systematic uncertainty of a factor of $1.6$--$2.7$ in $N_\mathrm{H}$ relative to the fixed ionization degree assumed in the original formulation. The total column mass is then \begin{equation}
    M_{\mathrm{res}} = \sum_{\mathrm{pixels}}
    \left(N_\mathrm{H} m_\mathrm{H} + 0.0851\, N_\mathrm{H} \times 3.97\, m_\mathrm{H}\right) \,d\, A_{\mathrm{pix}}
    \label{eq:resolved_mass}
\end{equation}
where $m_\mathrm{H}$ is the mass of the hydrogen atom, $\delta A$ is the projected pixel area, and $d$ is the line-of-sight depth derived from the three-dimensional reconstruction described in Section~\ref{sec:geometry_correction}.

\subsubsection{Disk-integrated (Sun-as-a-star) Mass}
\label{sec:star_mass}

For the stellar analog approach, we follow the methodology of \citet{namekataProbableDetectionEruptive2022}. We first compute a theoretical equivalent width ($\mathrm{EW}_{\mathrm{model}}$) corresponding to a filament with the same mean cloud model parameters ($\bar{S}$, $\bar{\tau}_0$, $\bar{v}_D$, $\bar{W}$) uniformly covering the entire stellar disk \citep{meinDifferentialCloudModels1988}:
\begin{equation}
    \mathrm{EW}_{\mathrm{model}} = \int_{\lambda} 
    \frac{\bar{S} - I_{0}(\lambda)}{I_{0,\mathrm{cont}}} 
    \left(1 - e^{-\tau(\lambda)}\right) d\lambda,
\end{equation}
where $\tau(\lambda)$ follows the Gaussian profile in Equation~\eqref{eq:tau} evaluated at the mean parameters. The disk filling factor $f$ is then derived by normalizing the observed equivalent width from the Sun-as-a-star synthesis (see Section~\ref{sec:sun_as_star}) by this theoretical maximum:
\begin{equation}
    f = \frac{\Delta \mathrm{EW}_{\mathrm{obs}}}{\mathrm{EW}_{\mathrm{model}}}.
\end{equation}
Finally, the disk-integrated mass $M_{\mathrm{total}}$ is estimated as 
\begin{equation}
    M_{\mathrm{total}} = \bar{m} \cdot f \cdot A_{\mathrm{disk}},
\end{equation}
where $A_{\mathrm{disk}}$ is the projected area of the full disk and $\bar{m}$ is the mean mass column density derived from the spatially averaged cloud model parameters using the same non-LTE ionization relations as in 
Section~\ref{sec:resolved_mass}.

\section{Results}
\label{sec:results}

This section presents statistical results regarding 1346 solar filaments we identified. We first examine the distributions of optical parameters derived from the cloud model inversion and the geometric properties measured from the segmentation masks. We then investigate projection effects on filament geometry to derive an empirical line-of-sight depth correction. Finally, we evaluate the filament mass estimates and quantify the impact of the depth correction on the consistency between the spatially resolved and Sun-as-a-star methods.

\subsection{Statistics of Optical and Physical Properties}
\label{sec:stats_optical}

\begin{figure}[t!]
    \centering
    \includegraphics[width=\columnwidth]{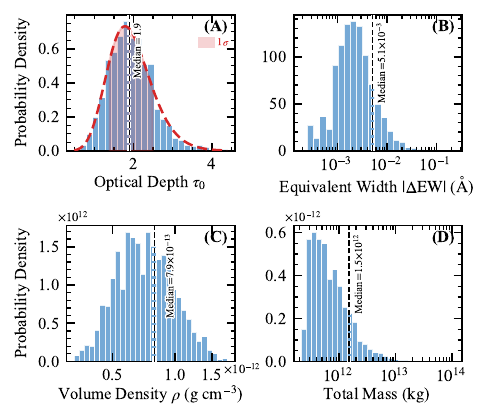}
    \caption{Distributions of physical parameters derived from the Sun-as-a-star cloud-model fitting for the full filament sample. (A) Optical depth $\tau_0$; (B) absolute equivalent width $|EW|$ in H$\alpha$; 
    (C) volume density $\rho$; (D) total mass. The dashed vertical line in each panel marks the median of the distribution. In panel (A), the red dashed curve shows the best-fit log-normal distribution selected by the BIC, and the shaded region indicates $1\sigma$ interval $[\tau_0 = 1.39,\ 2.50]$. }The y-axis in all panels represents the normalized probability density.
    \label{fig:optical_stats}
\end{figure}

Figure~\ref{fig:optical_stats} presents the statistical distributions of the optical and physical parameters derived from the cloud model inversion for the full sample.

The distribution of the line-center optical depth ($\tau_0$), as shown in Figure~\ref{fig:optical_stats}A, is right-skewed, and peaks at $\tau_0 \approx 2$. The Bayesian Information Criterion (BIC) analysis identifies the log-normal as the best-fit functional form with a median of $\tau_0 = 1.89$ and a $1\sigma$ interval of $[1.39,\ 2.50]$. The extended tail toward larger values indicates the presence of a minority of more optically thick cases. The majority of filaments are concentrated in the range $1 < \tau_0 < 3$, consistent with solar filaments being optically thick at the H$\alpha$ line center.

Figure~\ref{fig:optical_stats}B displays the distribution of the absolute equivalent width ($|\Delta \mathrm{EW}|$) derived from the Sun-as-a-star difference spectra. The distribution is right-skewed on a logarithmic scale, peaking near $10^{-3}$--$10^{-2}\ \mathrm{\AA}$ with a tail extending toward larger values, and a median of $5.1 \times 10^{-3}\ \mathrm{\AA}$. This reflects the inherently weak disk-integrated absorption signature produced by filaments, whose small fractional coverage of the solar disk makes it difficult to detect their spectral signature in unresolved stellar contexts.

The volume mass density $\rho$, shown in Figure~\ref{fig:optical_stats}C, follows a well-defined log-normal distribution with a median of $8.0 \times 10^{-13}\,\mathrm{g~cm^{-3}}$. Such a distribution is consistent with the typical range reported for quiescent prominence plasma \citep{labrossePhysicsSolarProminences2010}.

The total mass distribution (Figure~\ref{fig:optical_stats}D) also follows an approximately log-normal profile, but with a pronounced tail extending to large values. The median is $1.5 \times 10^{12} \ \mathrm{kg}$, while the distribution spans nearly two orders of magnitude from $\sim 10^{11}$ to $\sim 10^{13} \ \mathrm{kg}$. For reference, typical CME masses fall within $10^{11}-10^{13} \ \mathrm{kg}$ \citep{vourlidasComprehensiveAnalysisCoronal2010}, indicating that the most massive filaments in our sample carry a mass comparable to that of typical CME ejecta, thus serving as a major mass reservoir in solar eruptions.

\subsection{Geometric Reconstruction and Depth Correction}
\label{sec:geometry_correction}

\begin{figure*}[t!]
    \centering
    \includegraphics[width=\textwidth]{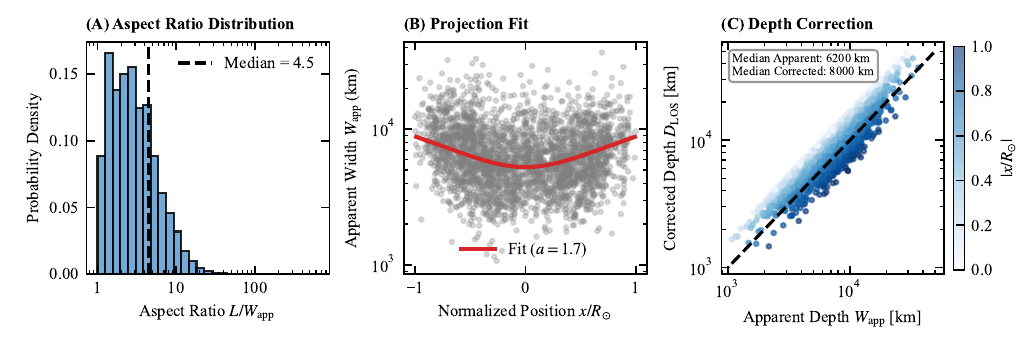}
    \caption{Geometric statistics and depth correction. (A) Distribution of the filament aspect ratio ($L/W_{\mathrm{app}}$). (B) Apparent width $W_{\mathrm{app}}$ as a function of normalized heliocentric position $x/R_\odot$. The red curve shows the best-fit projection model (Equation~\eqref{eq:depth_fitting}), yielding a depth-to-width ratio of $a = 1.7$. (C) Comparison between the apparent depth (assumed equal to $W_{\mathrm{app}}$) and the geometrically corrected depth $D_{\mathrm{LOS}}$, color encoded by heliocentric distance $|x/R_\odot|$. The dashed line indicates the one-to-one correspondence.}
    \label{fig:geometry_stats}
\end{figure*}

A critical challenge in estimating filament mass—both for solar and stellar observations—is the unknown geometric depth along the line of sight ($D_{\mathrm{LOS}}$). Our large-sample statistics provide an empirical validation of this parameter through projection effects across the solar disk.

Figure~\ref{fig:geometry_stats}A shows the distribution of the filament aspect ratio (projected length $L$ to apparent width $W_{\mathrm{app}}$), with a median of $4.5$, confirming the well-known elongated morphology of solar filaments.

To decouple the intrinsic transverse width ($W_0$) from the line-of-sight depth ($D_0$), we analyze the variation of the apparent width $W_{\mathrm{app}}$ as a function of the normalized heliocentric position $x/R_\odot$ (Figure~\ref{fig:geometry_stats}B). Near the disk center, the line of sight is nearly perpendicular to the filament spine, so only the transverse width contributes to $W_{\mathrm{app}}$. Toward the limb, the line of sight becomes increasingly perpendicular to the depth dimension, causing $W_{\mathrm{app}}$ to grow. We quantify this behavior by modeling the apparent width as a function of the heliocentric angle $\theta$ (where $\sin\theta = x/R_\odot$), assuming an elliptical cross-section with a depth-to-width ratio $a$:
\begin{equation}
    W_{\mathrm{app}}(\theta) = W_{0} \sqrt{\cos^2\theta + a^2 \sin^2\theta},
    \label{eq:depth_fitting}
\end{equation}
where $W_{0}$ is the characteristic transverse width and $a = D_{0}/W_{0}$ is the depth-to-width ratio. Fitting this model to the observed $W_{\mathrm{app}}$--$x/R_\odot$ trend yields $a \approx 1.68$, indicating that the typical filament depth exceeds its transverse width. The substantial scatter around the fitted relation reflects the intrinsic diversity of filament morphology, including variations in physical size, aspect ratio, and other projection effects. Therefore, the inferred relation should be interpreted in a statistical sense and is not intended to provide a precise geometric description for individual filaments. Based on such a relation, we infer the statistical three-dimensional morphology of the filament population: the characteristic length-to-width-to-depth ratio is approximately $4.5 : 1 : 1.7$.

Applying this geometric correction to each filament, we derive the corrected line-of-sight depth $D_{\mathrm{LOS}}$. As shown in Figure~\ref{fig:geometry_stats}C, the median depth increases from an apparent value of $6200\,\mathrm{km}$ to a corrected value of $8000\,\mathrm{km}$, with the correction being largest for filaments located near the limb (largest $|x/R_\odot|$), as expected from projection geometry.

\subsection{Scaling Relations of Filament Mass}
\label{sec:scaling}

\begin{figure*}[t!]
    \centering
    \includegraphics[width=\textwidth]{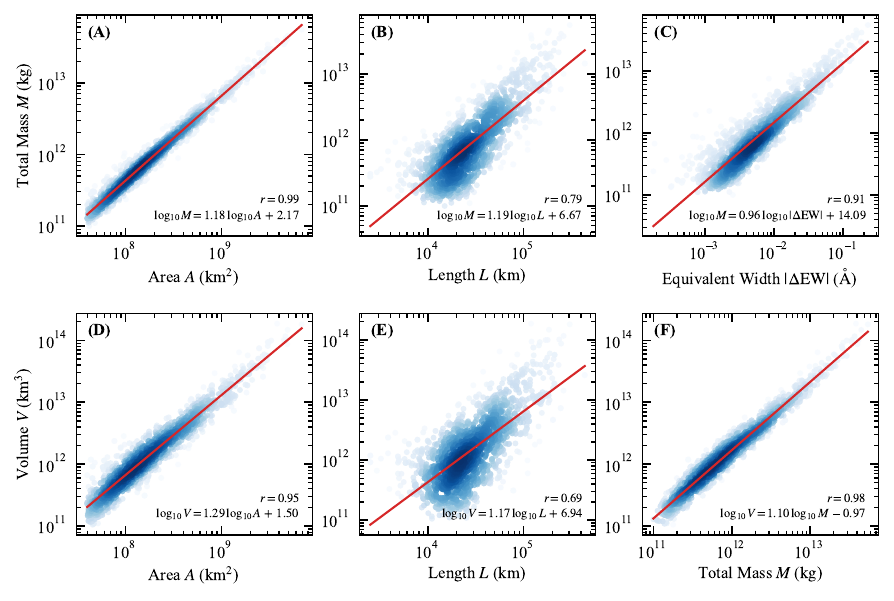}
    \caption{Scaling relations among the paprameters of the filaments in log--log space. Point colors encode the local number density estimated via kernel density estimation, with darker blue indicating higher concentration. The red solid lines show the least-squares linear fits in log--log space, with the best-fit relations and Pearson correlation coefficients $r$ given in the lower-right corners. Panels (A)--(C) present the scaling relations between total mass $M$ and area $A$ (A), length $L$ (B), and equivalent width $|\Delta\mathrm{EW}|$ (C). Panels (D)--(F) show the geometric scaling relations between the derived volume $V$ and area $A$ (D), length $L$ (E), and total mass $M$ (F).}
    \label{fig:mass_correlations}
\end{figure*}

Panels (A)--(C) of Figure~\ref{fig:mass_correlations} presents the empirical scaling relations between the total filament mass and four key physical parameters in log–log space. We adopt log--log linear fits for all relations, as this form provides a directly interpretable, scale-free relationship that remains stable under extrapolation beyond the observed data range, and adequately captures the dominant trend relative to the intrinsic scatter of the data.

The strongest correlation is with the projected area $A$ Figure~\ref{fig:mass_correlations}A, yielding a Pearson coefficient of $r = 0.99$ and a log-space regression slope of $k = 1.2$, consistent with a near-linear relationship in log--log space. A positive but weaker correlation is observed with the projected length $L$ (Figure~\ref{fig:mass_correlations}B, $r = 0.79$). Comparing our mass--length relation with the statistical survey of filament eruptions by \citet{kotaniUnifiedRelationshipCold2023a} (Figure~10) reveals a systematic offset: at comparable filament lengths, our mass estimates are 2--3 orders of magnitude higher than those reported by \citet{kotaniUnifiedRelationshipCold2023a}. This discrepancy is attributable to the difference in detection methodology: \citet{kotaniUnifiedRelationshipCold2023a} identified ejections based on blue-shifted velocity signatures, effectively counting only kinematically active plasma, whereas our method captures the entire filament structure, including stationary bulk plasma, thereby covering the total mass more completely.

Crucially for stellar applications, Figure~\ref{fig:mass_correlations}C reveals a strong correlation ($r = 0.91$) between the total mass and the Sun-as-a-star equivalent width $|\Delta\mathrm{EW}|$ in log--log space, with a near-linear scaling index of $k = 0.96$. This indicates that the disk-integrated $|\Delta\mathrm{EW}|$ can serve as a robust proxy for the total filament mass, even in the absence of spatial resolution.

\begin{figure*}[t!]
    \centering
    \includegraphics[width=\textwidth]{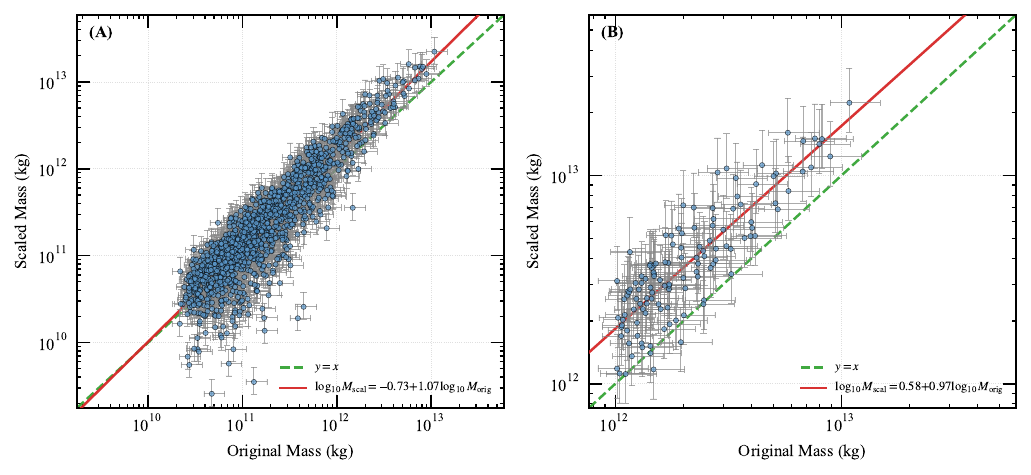}
    \caption{Comparison between the spatially resolved mass ($M_{\mathrm{orig}}$) and the Sun-as-a-star estimated mass ($M_{\mathrm{scal}}$). The red solid line shows the linear fit in logarithmic space, and the green dashed line indicates the one-to-one relation ($y = x$). (A) Full sample spanning more than two orders of magnitude ($10^{10} - 10^{13}$\,kg), with best-fit relation $\log_{10}(M_{\rm scal}) = 1.07\log_{10}(M_{\rm orig}) - 0.73$ and the coefficient of determination $R^2 = 0.82$. (B) High-mass subsample restricted to $M_{\rm orig},\,M_{\rm scal} > 10^{12}$\,kg, with best-fit relation $\log_{10}M_{\rm scal} = 0.97\log_{10}M_{\rm orig} + 0.30$, showing that $M_{\rm scal}$ overestimates $M_{\rm orig}$ by a roughly constant factor across the high-mass regime.}
    \label{fig:mass_comparision}
\end{figure*}

\subsection{Sub-Euclidean scaling and evidence for a porous, sheet-like structure}
\label{sec:fractal}

Panels (D)--(F) of Figure~\ref{fig:mass_correlations} show the scaling relations between the derived volume $V$ and the characteristic parameters of filaments. The best-fit power laws are

\begin{equation}
    V\propto A^{1.29},\qquad
    V\propto L^{1.17},\qquad
    V\propto M^{1.10}.
    \label{eq:vscaling}
\end{equation}

For a compact, homogeneous three-dimensional structure with self-similar growth, the expected relations are $V\propto A^{3/2}$ and $V\propto L^3$, respectively. However, the fitted power-law indices are systematically smaller than the Euclidean expectations, indicating a sub-Euclidean scaling behavior. The volume--area relation ($r=0.95$) corresponds to an effective scaling dimension of $D_{\rm eff}\simeq2\times1.29\approx2.6$, suggesting that filament growth does not involve proportional expansion in all three dimensions. Instead, because the volume is calculated as $V=A\,D$ with the line-of-sight depth $D$ scaled from the observed width, the flatter volume--length relation indicates that filaments preferentially extend along their length while remaining relatively thin.

The volume--mass relation in Figure~\ref{fig:mass_correlations}F provides an additional constraint on the filling factor of cool filament material within the inferred volume. The mass is independently derived by summing the pixel-by-pixel column densities over the filament area and does not rely on the assumed geometrical volume. The tight correlation $V\propto M^{1.10}$ imply
\begin{equation}
    \bar{\rho}=\frac{M}{V}\propto M^{-0.1},
    \label{eq:rho}
\end{equation}
indicating that the mean density decreases with filament size. That is to say, for larger filaments, cool materials occupy an increasingly smaller fraction of their volume.

Together, the sub-Euclidean geometric scaling and the decreasing mean density suggest that filaments are not compact, uniformly filled three-dimensional structures, but instead possess a porous and highly structured internal structure. This interpretation is consistent with high-resolution observations showing that quiescent filaments/prominences consist of numerous thin threads separated by voids and cavities \citep[e.g.,][]{Berger2008,Berger2010,Gibson2018}. As filaments grow, these fine structures do not fill the additional volume proportionally, likely producing the observed deviations from Euclidean scaling.

\subsection{Validation of the Sun-as-a-star Mass Estimation}
\label{sec:mass_evaluation}

Figure~\ref{fig:mass_comparision} compares the Sun-as-a-star estimated mass $M_{\rm scal}$ with the pixel-by-pixel resolved intrinsic mass $M_{\rm orig}$. Panel~(A) presents the full sample, spanning more than two orders of magnitude ($10^{10} - 10^{13} \ \mathrm{kg}$). The two approaches show a high degree of consistency ($R^2 = 0.82$), with a best-fit scaling relation of:
\begin{equation}
    \log_{10} (M_{\rm scal}) = 1.07\,\log_{10} (M_{\rm orig}) - 0.77.
    \label{eq:scaling_full}
\end{equation}

A key finding arises from comparing the scaling behavior across different mass regimes. For high-mass filaments ($> 10^{12} \ \mathrm{kg}$; Panel B), the fitting slope is remarkably close to unity ($k = 0.97$), demonstrating that the Sun-as-a-star method is highly reliable in recovering the true mass of filaments and its variation for large events. However, the best fit line consistently appears to be located above the $y = x$ relation, indicating a tendency to systematically overestimate the mass of large filaments by a nearly constant factor. 

In contrast, the behavior in the low-mass regime is more complex. While a formal fit for the lowest-mass subset is less constrained, visual inspection of the scatter plot reveals a non-negligible population of small-scale filaments whose $M_{\rm scal}$ is significantly lower than $M_{\rm orig}$. This suggests that the Sun-as-a-star approach may be prone to underestimating the mass of smaller filaments. This localized underestimation at the low-mass end explains why the full-sample slope ($1.07$ in Panel A) is notably steeper than the high-mass slope ($0.97$ in Panel B); the "downward pull" exerted by these underestimated low-mass samples effectively induces a pivot in the global regression linear relation.

\section{Discussion}
\label{sec:discussion}

\subsection{Physical Interpretation of Filament 3D Morphology}
\label{sec:discuss_geometry}

The statistical 3D morphology derived from our projection analysis yields a characteristic length-width-depth ratio of $4.5:1:1.7$ for quiescent filaments. The elongated morphology ($L/W \approx 4.5$) is consistent with filament formation along PILs, as the long axis is known to align with the underlying magnetic channel \citep{martinConditionsFormationMaintenance1998}.

More significantly, the depth-to-width ratio $a \approx 1.7$ challenges the common assumption in stellar studies that the geometric depth equals the width \citep{namekataProbableDetectionEruptive2022}, suggesting that the filament plasma extends over a coronal depth greater than its transverse width. The adoption of $a = 1$ therefore leads to an underestimate of the line-of-sight depth by up to $\sim\!30\%$, which gives a proportional underestimate in the volume and mass of the filament. Applying the solar-calibrated ratio to the mass estimation of \citet{namekataProbableDetectionEruptive2022} increases the inferred mass by a factor of $\sim\!1.58$, yielding a revised estimate of $3.43 \times 10^{18}$\,g. As a population-level calibration for stellar applications, where individual geometry is always inaccessible, this empirical constraint represents a significant improvement over prior ad hoc assumptions.

\subsection{Comparison with Previous Mass Measurements}
\label{sec:discuss_comparison}

The median filament mass of $7.2 \times 10^{11}$\,kg is broadly consistent with previous estimates: \citet{gilbertCOMPARINGSPATIALDISTRIBUTIONS2010} reported prominence masses in the range $10^{10}$--$10^{11}$\,kg from EUV opacity measurements of nine prominences. For comparison, our derived mean density of $\sim10^{-12}$\,g\,cm$^{-3}$ corresponds to a hydrogen number density of $n_H \sim 6\times10^{11}$\,cm$^{-3}$, which is consistent in order of magnitude with the characteristic value of $\sim10^{11}$\,cm$^{-3}$ adopted for quiescent prominences at temperatures of 6000--8000\,K \citep{heinzelHinodeTRACESOHO2008}.

A direct comparison with \citet{kotaniUnifiedRelationshipCold2023a} is instructive. Although both studies employ cloud model fitting to H$\alpha$ spectra, \citet{kotaniUnifiedRelationshipCold2023a} targeted the kinematically active ejecta during the eruptions, recovering only the momentum-carrying plasma component. Our study, in contrast, integrates the full column density of quiescent filaments regardless of line-of-sight velocity, providing a census of the total pre-eruption mass reservoir. The systematically higher masses we derive are therefore physically expected, as the quiescent filament mass necessarily exceeds the fraction ejected during the eruption.

The most massive filaments extending over $10^{13}\,\mathrm{kg}$ are comparable to typical CME masses \citep{webbCoronalMassEjections2012a}, suggesting that the most massive filaments in our sample represent viable progenitors of large-scale eruptions.

\subsection{Uncertainties and Limitations}
\label{sec:discuss_uncertainties}

The uncertainties discussed in Section~\ref{sec:scaling} are purely statistical in nature, describing how the uncertainties in the fitted model parameters propagate across the full sample. Several additional sources of uncertainty are not captured by this framework and warrant explicit discussion.

The spatially resolved mass $M_{\mathrm{orig}}$ is computed assuming that the line-of-sight depth $d$ of each filament is accurately described by the empirical depth-to-width relation. In the absence of a direct observational method to constrain the three-dimensional geometry of individual filaments, $d$ is treated as a fixed, error-free quantity in the present analysis. This assumption is justified at a statistical level since the mean ratio $\bar{a} = 1.7$ is well constrained by the full sample; however, individual filaments inevitably deviate from this mean geometry. Incorporating the per-event geometric uncertainty would require an independent depth estimate for each filament, which is not available with current single-viewpoint observations. If such individual geometric diversities are propagated into $M_{\mathrm{orig}}$, the dispersion in Figure~\ref{fig:mass_comparision} will increase substantially, and the present $\log_{10} \sigma_\Delta = 0.26$ should therefore be regarded as a lower bound of the true event-level uncertainty.

All uncertainties quantified in this work, including both Monte Carlo error bars per-event and the residual scatter $\sigma_\Delta$, are statistical quantities derived from the population distribution of the spectral and geometric parameters. They characterize how well the framework performs on average across a large sample but do not reflect the full uncertainties applicable to any single filament event. For an individual filament whose true optical depth, line width, or geometry departs significantly from the population mean, the actual mass error could be considerably larger than those suggested by the error bars in Figure~\ref{fig:mass_comparision}. The Sun-as-a-star approach is therefore most reliable when applied to ensemble-averaged or statistically aggregated quantities, and its single-event mass estimates should be interpreted with appropriate caution.

\subsection{Implications for Stellar Filament Mass Estimation}
\label{sec:discuss_stellar}

The near-unity power-law slope ($k = 0.96$) of the $M-|\Delta\mathrm{EW}|$ relation demonstrates that the disk-integrated equivalent width is approximately a linear proxy for total filament mass. This linearity arises physically because optically thin H$\alpha$ absorption is proportional to column density, which scales linearly with total mass for a fixed area.

The empirical scaling relation,
\begin{equation}
    \log_{10} (M_{\mathrm{true}}) = \frac{\log_{10} (M_{\mathrm{star}}) + 0.73}{1.07}
    \label{eq:scaling_stellar}
\end{equation}
provides a direct correction for stellar filament mass estimates obtained via the \citet{namekataProbableDetectionEruptive2022} methodology: after computing $M_\mathrm{star}$ as described in Section~\ref{sec:mass_evaluation}, we can then invert Equation~\eqref{eq:scaling_stellar} to obtain $M_\mathrm{true}$.

This relation constitutes the most empirically constrained baseline presently available for diagnosing stellar filament and coronal mass ejection (CME) masses from disk-integrated spectroscopy.

\section{Conclusion}
\label{sec:conclusion}

We present the first large-sample Sun-as-a-star study of solar filaments with CHASE H$\alpha$ hyperspectral observations. A total of 1,346 quiescent filaments are statistically analyzed, with the intention of characterizing filament mass and geometry and calibrating the Sun-as-a-star mass estimation technique for stellar applications. Our main conclusions are as follows.

\begin{enumerate}

\item The line-center optical depth follows a right-skewed distribution with a median of $\tau_0=1.9$. The total mass spans two orders of magnitude, with a median of $1.5 \times 10^{12}\,\rm{kg}$, consistent with previous EUV and H$\alpha$ estimates. The most massive filaments extend beyond $10^{13}\,\rm{kg}$, comparable to typical masses of major CMEs.

\item The analysis of projection-dependent apparent width yields a characteristic 3D morphology of $L:W:D = 4.5:1:1.7$. The depth-to-width ratio $a \approx 1.7$ implies that simply assuming an equal depth and width with $a = 1$ could severely underestimate the line-of-sight depth, and hence the mass, by up to $\sim\!30\%$. The corrected median depth is $\sim\!8000\,\rm{km}$.

\item The masses of filaments correlate most strongly with their projected areas, following $M \propto A^{1.18}$ in log--log space, indicating that the mass is nearly proportional to the projected area. A near-linear correlation between the masses and the Sun-as-a-star equivalent widths, following $M \propto |\Delta\mathrm{EW}|^{0.96}$, enables the use of $|\Delta \mathrm{EW}|$ as an effective mass proxy for inferring filament masses in spatially-unresolved stellar observations.

\item For filament masses, by incorporating the geometric constraints derived from our statistical analysis, the Sun-as-a-star framework achieves an $R^2 = 0.82$ agreement with the spatially resolved intrinsic masses over two orders of magnitude, yielding the empirical calibration of equation~\eqref{eq:scaling_stellar}. This provides a physically reliable tool for estimating stellar filament and CME masses from disk-integrated H$\alpha$ spectra.

\end{enumerate}
 
\begin{acknowledgments}

This work is supported by the National Natural Science Foundation of China under grant 12525305, 12333009, and the Fundamental Research Funds for the Central Universities (KG202506). 
Observation data is from the CHASE mission supported by the China National Space Administration. The calculation was done on the computing facilities in High Performance Computing Center of Nanjing University. 

\end{acknowledgments}

\bibliography{sample/job1}{}
\bibliographystyle{aasjournalv7}

\end{document}